\documentclass[useAMS,usenatbib,usegraphicx]{mn2e}
\usepackage{epsfig}
\usepackage{amsmath} 
\usepackage{rotating}           
\usepackage{color}     
\usepackage{graphicx}
\usepackage{times}
\usepackage{upgreek} 

\def\kms{km ${\rm s}^{-1}$}

\def\Mo{M$_\odot$}

\def\ccm {$\hbox{{\rm cm}}^{-3}$}    
\def\scm  {$\hbox{{\rm cm}}^{-2}$}    
\def \AL {$\alpha $}     
\def \HI {H{\sc \,i}}

\def\dg{$^{\circ}$}
\def\lapp{\ifmmode\stackrel{<}{_{\sim}}\else$\stackrel{<}{_{\sim}}$\fi}
\def\gapp{\ifmmode\stackrel{>}{_{\sim}}\else$\stackrel{>}{_{\sim}}$\fi}

\title[DLA spin temperature and star formation history]{A possible connection between the spin temperature of  damped Lyman-\boldmath{\AL} absorption systems and star formation history}
\author[S. J. Curran et al.]{S. J. Curran\thanks{Stephen.Curran@vuw.ac.nz}\\
School of Chemical and Physical Sciences, Victoria University of Wellington, PO Box 600, Wellington 6140, New Zealand\\
}

\begin{document}

 \date{Accepted ---. Received ---; in original form ---}

\pagerange{\pageref{firstpage}--\pageref{lastpage}} \pubyear{2017}

\maketitle

\label{firstpage}
\begin{abstract}
  We present a comprehensive analysis of the spin temperature/covering factor degeneracy, $T_{\rm spin}/f$, in damped
  Lyman-\AL\ absorption systems. By normalising the upper limits and including these via a survival analysis, there is,
  as previously claimed, an apparent increase in $T_{\rm spin}/f$ with redshift at $z_{\rm abs}\gapp1$. However, when we
  account for the geometry effects of an expanding Universe, neglected by the previous studies, this increase in $T_{\rm
    spin}$ at $z_{\rm abs}\gapp1$ is preceded by a decrease at $z_{\rm abs}\lapp1$.  Using high resolution radio images
  of the background continuum sources, we can transform the $T_{\rm spin}/f$ degeneracy to $T_{\rm spin}/d_{\rm
    abs}^{~~~2}$, where $d_{\rm abs}$ is the projected linear size of the absorber. Again, there is no overall increase
  with redshift, although a dip at $z_{\rm abs}\approx2$ persists. Furthermore, we find $d_{\rm abs}^{~~~2}/T_{\rm
    spin}$ to follow a similar variation with redshift as the star formation rate, $\psi_{*}$. This suggests that,
  although the total hydrogen column density, $N_{\rm HI}$, shows little relation to $\psi_{*}$, the fraction of the
  cold neutral medium, $\int\!\tau_{\rm obs}\,dv/N_{\rm HI}$, may. Therefore, further efforts to link the neutral gas
  with the star formation history should also consider the cool component of the gas.
\end{abstract}
 
\begin{keywords}
galaxies: high redshift --  galaxies: star formation  -- galaxies: evolution -- galaxies: ISM -- quasars: absorption lines --  radio lines: galaxies
\end{keywords}

\section{Introduction} 
\label{intro}
 
Studies of redshifted absorption systems lying along the sight-lines to distant quasi-stellar objects (QSOs) provide a
probe of the evolution of quiescent galaxies over the history of the Universe. Of particular interest are the so called
damped Lyman-$\alpha$ absorption systems (DLAs), since their high defining neutral hydrogen (\HI) column densities of
$N_{\rm HI}\ge2\times10^{20}$ \scm\ act as signatures for distant gas rich galaxies\footnote{DLAs may account or at
  least 80\% of the neutral gas mass density in the Universe \citep{phw05}.}, fuelling the possible star formation
\citep{nkp+17}.\footnote{From two $z\sim4$ DLAs recently detected in [C{\sc \,ii}] 158~$\mu$m emission. This arises from heated
  dust, generally believed to be caused by a  stellar population (e.g. \citealt{mcc+05}), although heating by an active nucleus is
  also a possibility \citep{cur09}.}  The Lyman-$\alpha$ transition occurs in the ultra-violet band at $\lambda_{\rm
  rest}= 1216$ \AA, and so is only redshifted into the atmospheric observing window at $z\gapp1.7$. In addition to
space-based observations of this transition, neutral hydrogen can be detected from $z\geq0$ via the $\lambda_{\rm
  rest}=21.1$-cm spin-flip transition.  The combination of \HI\ 21-cm and Lyman-\AL\ data can yield important
information about the high redshift Universe, such as testing for any evolution in the fundamental constants (see
\citealt{cdk04,twm+05}) and measuring the temperature, and thus star-forming potential, of the neutral gas in the most
distant galaxies (e.g. \citealt{wgp03}).

The spin temperature, $T_{\rm spin}$, of the gas is a measure of the population of the lower hyperfine level ($F=1$),
where the gas can absorb 21-cm photons, relative to the upper hyperfine level ($F=2$). 
Thus, the spin temperature can be raised via excitation to the upper hyperfine level by 21-cm absorption
\citep{pf56}, excitation above ground state by Lyman-\AL\ absorption \citep{fie59} and collisional excitation
\citep{be69}. Presuming that the Lyman-\AL\ and 21-cm absorption trace the same sight-line, the spin
temperature can be obtained by comparing the velocity integrated optical depth of the 21-cm absorption strength,
$\int\!\tau\,dv$, with the total neutral hydrogen column density, via (e.g. \citealt{rw00})
\begin{equation}
N_{\rm HI}  =1.823\times10^{18}\,T_{\rm  spin}\int\!\tau\,dv.
\label{enew_full}
\end{equation}
The observed optical depth is the ratio of the line depth, $\Delta S$, to the observed background flux, $S_{\rm obs}$, and is
related to the intrinsic optical depth via
\begin{equation}
\tau \equiv-\ln\left(1-\frac{\tau_{\rm obs}}{f}\right) \approx  \frac{\tau_{\rm obs}}{f}, {\rm ~for~}  \tau_{\rm obs}\equiv\frac{\Delta S}{S_{\rm obs}}\lapp0.3,
\label{tau_obs}
\end{equation}
where the covering factor, $f$, is the fraction of $S_{\rm obs}$ intercepted by the absorber.
Therefore, in the optically thin regime (where $\tau_{\rm obs}\lapp0.3$), Equ. \ref{enew_full} can be rewritten as
\begin{equation}
N_{\rm HI}  \approx 1.823\times10^{18}\,\frac{T_{\rm  spin}}{f}\int\!\tau_{\rm obs}\,dv, 
\label{enew}
\end{equation}
so that the comparison of the neutral hydrogen column density with the observed velocity integrated optical depth of the
21-cm absorption gives the degenerate $T_{\rm spin}/f$.  Since, by definition, $\tau_{\rm obs}< f\leq1$ (the lower limit
being imposed by Equ.~\ref{tau_obs}, \citealt{obg94}), then $\tau_{\rm obs}\leq\tau$ and if $\tau_{\rm obs} < \tau$, the
observed absorption strength is diluted. That is, assuming $f=1$, when in reality $f<1$, has the effect of artificially raising the
apparent spin temperature (see figure 1 of \citealt{cmp+03}).

Since \citet{kc02}  noted a mix of 21-cm detections and non-detections in DLAs at $z_{\rm abs}\lapp2$, whereas at $z_{\rm
  abs}\gapp2$ detections were rare, they suggested that 
the lower redshift DLAs had a mix of spin temperatures with the higher redshift DLAs having exclusively high spin
temperatures, suggesting an ``evolution'' in the temperature of the gas.  However, \citet{cmp+03} argued that,
since $T_{\rm spin}/f$ was degenerate, the values of $T_{\rm spin}$ obtained by \citet{kc02} could be due to their
estimates of $f=1$ for all of the highest redshift absorbers. 
Following this, \citet{cw06} showed that, due to the geometry of an expanding Universe, low redshift absorbers
could be at lower angular diameter distances than the background QSOs, whereas at higher redshift the angular 
diameter distance to the
absorber is always similar to that to the QSO ($DA_{\rm DLA}\approx DA_{\rm QSO},\, \forall \,z_{\rm abs} \gapp1.6$), no
matter the relative redshifts (see Sect. \ref{cfg}). Since, to a first order, the covering factor is proportional to the
ratio of the angular diameter distances, 
\citeauthor{cw06} argued that this distribution could in fact be the cause of the observed $T_{\rm spin}/f$---redshift
distribution, with \citet{cur12} showing that the factor of two between the mean values of $T_{\rm spin}/f$ ($1800$ K at
$z_{\rm abs}< 2$ and $3600$ K at $z_{\rm abs}>2$) could be accounted by the difference in angular diameter distance
ratios between these two redshift ranges. Thus, the observed $T_{\rm spin}/f$--redshift distribution could be explained
by the geometry effects of an expanding Universe, not requiring an evolution in the spin temperature.

Since then, several studies have reasserted that the spin temperature increases with redshift, although these avoid
the geometry effects by only considering a limited redshift range ($z_{\rm abs} > 2$ by \citealt{rmgn13} and
$z_{\rm abs} > 1$ by \citealt{kps+14}). 
 \begin{figure}
\centering \includegraphics[angle=-90,scale=0.50]{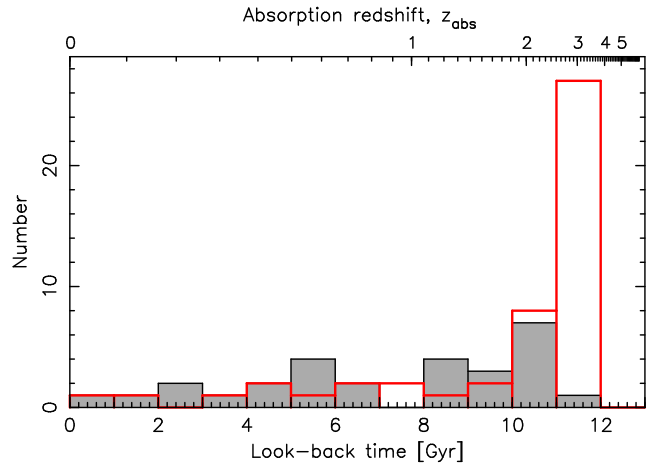}
\caption{The distribution of the 21-cm detections (shaded histogram) and non-detections (unshaded) with look-back time.
  Throughout the paper we use a standard $\Lambda$ cosmology with $H_{0}=71$~km~s$^{-1}$~Mpc$^{-1}$, $\Omega_{\rm
    matter}=0.27$ and $\Omega_{\Lambda}=0.73$.}
\label{LBT}
\end{figure} 
This excludes over half of the Universe (as well as 30\% of the sample, Fig.~\ref{LBT}), and obviously cannot test the original
hypothesis of \citet{kc02} that there is a difference in the spin temperatures between the low and high redshift
samples.  In this paper we undertake a full analysis by incorporating the geometry effects in order to determine whether
there is an evolution in the spin temperature
over the full $\approx12$ Gyr probed.

\section{Analysis}
\subsection{Normalisation and inclusion of the limits}

\begin{figure}
\centering \includegraphics[angle=-90,scale=0.50]{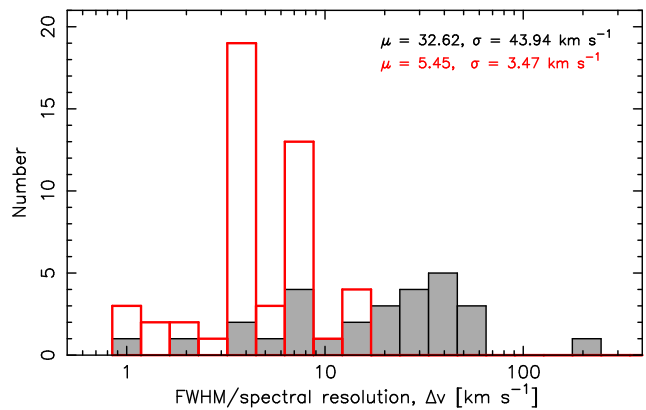}
\caption{The distribution of spectral resolution (for the non-detection, unshaded) and the line-widths (for the
  detections, shaded).  The non-detections span a range of $0.93 - 15$ \kms, which are too disparate to show clearly on a linear scale.}
\label{vel-hist}
\end{figure}

In order to normalise the search sensitivities, which are quoted over a large range of spectral resolutions
(Fig. \ref{vel-hist})\footnote{The 21-cm searches are compiled from
  \citet{dm78,bs79,bw83,ck00,kcsp01,kgc01,kpec09,kem+12,kps+14,bdv01,kc01a,kc02,cmp+03,ctp+07,ctd+09,ykep07,gsp+09,gsp+09a,ekp+12,sgp+12,rmgn13,kan14}.},
and thus meaningless to compare directly, we have to re-sample the r.m.s. noise levels to a common channel width.  We
then use this as the full-width half maximum (FWHM) of the putative absorption profile, thus giving a normalised
velocity integrated optical depth per channel limit to all of the non-detections (see \citealt{cur12} for details).
We use the mean $\left< {\rm FWHM}\right> = 33$ \kms\ for the detections to recalculate 
the limit to  $T_{\rm spin}/f$ for each non-detection. Note that there is
no significant difference between the FWHMs of the low and high redshift intervening 21-cm absorbers \citep{cdda16}, and
so we do not apply any redshift dependence.

\begin{figure}
\centering \includegraphics[angle=-90,scale=0.52]{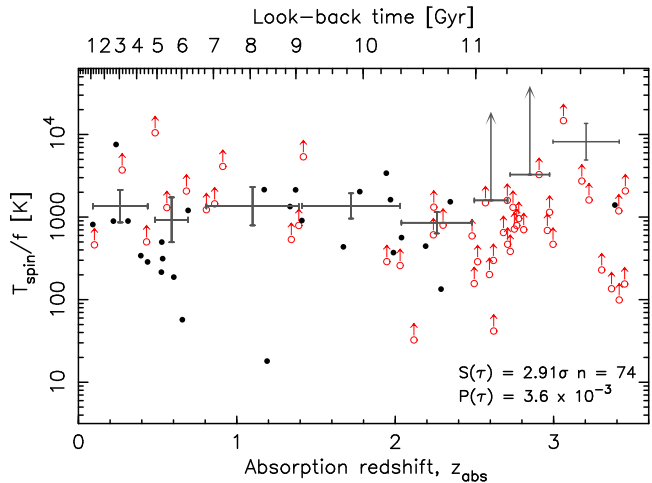}
\caption{The spin temperate/covering factor ratio versus the redshift for the DLAs searched in 21-cm absorption.  The
  filled circles show the detections, the unfilled circles  the lower limits and the error bars the
  binned values including the limits. The horizontal bars show the range of points in the bin and the vertical
  error bars the $1\sigma$ uncertainty in the mean value.}
\label{DLA-z}
\end{figure}

In order to fully utilise these data, we include the lower limits to $T_{\rm spin}/f$ (from the non-detections) via
the {\em Astronomy SURVival Analysis} ({\sc asurv}) package \citep{ifn86}, which are added to the analysis
as censored data points (Fig. \ref{DLA-z}, top).  For the bivariate data, a generalised non-parametric Kendall-tau test gives a
probability of $P(\tau) = 3.60\times10^{-3}$ of the observed $T_{\rm spin}/f$--$z_{\rm abs}$ correlation arising
by chance, which is significant at $S(\tau) = 2.91\sigma$, assuming Gaussian statistics.
For the binned data, the limits are incorporated, via the Kaplan--Meier estimator, which gives
a maximum-likelihood estimate based upon the parent population \citep{fn85}.  As a visual representation to complement
the Kendall-tau test, it is apparent that $T_{\rm spin}/f$ does exhibit an increase with redshift.

\subsection{Correcting for geometry} 
\label{cfg}

\begin{figure} 
\centering \includegraphics[angle=0,scale=0.42]{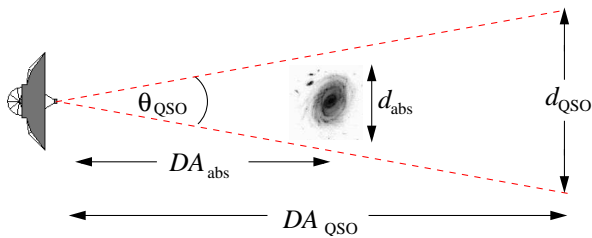} 
\caption{The projected absorber ($d_{\rm abs}$) and emitter cross-sections ($d_{\rm QSO}$) with respect to their angular
  diameter distances ($DA_{\rm abs}$ and $DA_{\rm QSO}$).$\theta_{\rm QSO}$ is the angle
  subtended by the background emission.  For a disk inclination of $i$ (where $i=90$\dg\ for face-on) $d_{\rm abs} =
  d_{\rm HI}\sin i$, where $d_{\rm HI}$ is the actual diameter of the \HI\ disk. For intervening absorption we expect a
  close to face-on inclination (e.g. \citealt{cras16}). Adapted from \citet{cur12}.}
\label{dla_schem}
\end{figure}

\begin{figure}
\centering \includegraphics[angle=-90,scale=0.52]{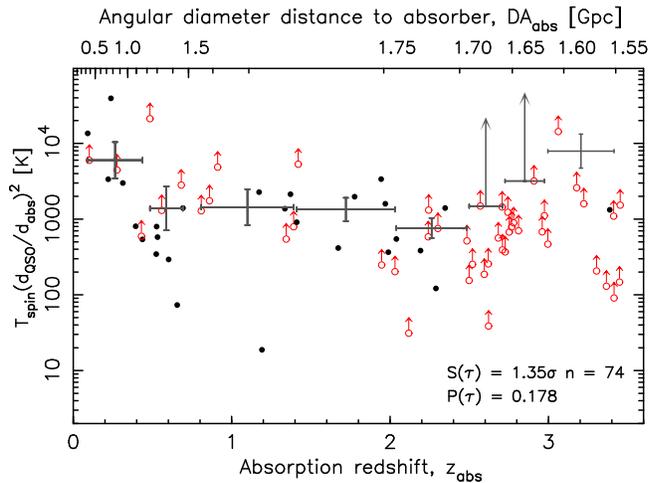}
\caption{The  $T_{\rm spin}/f$--$z$ distribution (Fig. \ref{DLA-z}) corrected for the angular diameter distances.}
\label{DLA-corr}
\end{figure}

The positive $T_{\rm spin}/f$--$z_{\rm abs}$ correlation could support the recent studies which conclude that there is
an increase in the spin temperature with redshift. However, as stated in Sect. \ref{intro}, none of these account for
the geometry effects of an expanding Universe, thus neglecting half of its history, which must be 
accounted for prior to invoking any cosmic evolution.  In the small angle approximation,
the covering factor is given by (\citealt{cur12,azdc16}, see Fig. \ref{dla_schem}).
\begin{equation}
f= \left\{   
\begin{array}{l l}
\frac{d_{\rm abs}^{~~~2}}{DA_{\rm abs}^{~~~2}.\theta_{\rm QSO}^{~~~~~2}} & \text{ if } \theta_{\rm abs} < \theta_{\rm QSO}\\
  1  & \text{ if } \theta_{\rm abs} \geq\theta_{\rm QSO},\\
\end{array}
\right.  
\label{f}
\end{equation}
where the angular diameter distance to a source is given by
\begin{equation}
DA = \frac{DC}{z+1},{\rm ~where~} DC = \frac{c}{H_0}\int_{0}^{z}\frac{dz}{H_{\rm z}/H_0} 
\label{equ:DA}
\end{equation}
is the line-of-sight co-moving distance (e.g. \citealt{pea99}), 
in which $c$ is the speed of light, $H_0$ the Hubble constant, $H_{\rm z}$ the Hubble parameter at redshift $z$ and 
\begin{equation}
\frac{H_{\rm z}}{H_{0}} = \sqrt{\Omega_{\rm m}\,(z+1)^3 + (1-\Omega_{\rm m} - \Omega_{\Lambda})\,(z+1)^2 + \Omega_{\Lambda}}.
\end{equation}
For a standard $\Lambda$ cosmology, this gives a peak in the angular diameter distance at $z\approx1.6$, meaning
that objects beyond this redshift are all essentially at the same angular diameter distance of $DA\approx1500 - 1700$ Mpc,
irrespective of their co-moving distances.\footnote{See the top scale of Fig. \ref{DLA-corr}.} This has the consequence
that at low redshift it is possible for $DA_{\rm abs} < DA_{\rm QSO}$ (when $z_{\rm abs} < z_{\rm QSO}$), giving the
familiar angular size--distance relation. However, at high
redshift only $DA_{\rm abs} \approx DA_{\rm QSO}$ is possible, which will introduce a bias to the covering
factor between the low and high redshift regimes.
Therefore, in Fig. \ref {DLA-corr} we correct for the angular diameter distances by scaling $T_{\rm spin}/f$ by
$\left(\frac{\,DA_{\rm QSO}}{DA_{\rm abs}}\right)^{2}$, giving $T_{\rm spin}\left(\frac{d_{\rm QSO}}{d_{\rm
      abs}}\right)^2$ on the ordinate. This causes the correlation evident in Fig.~\ref{DLA-z} to
disappear, although there will appear to be a strong increase in  $T_{\rm  spin}/f$  with redshift if the $z_{\rm abs}\lapp1$ data
are ignored \citep{rmgn13,kps+14}.
 
\subsection{Normalisation by emitter extent}
\label{bcfg}

The correction for the geometry effects indicates a decrease in $T_{\rm spin}/f$ with redshift over the most recent half
of the Universe's history, followed by an increase over the first half. Whether this is dominated by the spin
temperature or the covering factor remains unknown, although we can attempt to break the degeneracy by deconstructing
the covering factor.
Previous studies attempt to estimate the covering factor by assuming that this is given by the ratio of the flux from
the compact unresolved component of the radio emission to the total radio continuum flux (e.g. \citealt{bw83,
  klm+09,kem+12,kps+14}). However, the ratio of the fluxes contains no information on the depth of the line when
the extended continuum emission is resolved out, the extent of the absorber, nor how this is aligned along the
sight-line to the QSO.

\begin{figure}
\centering \includegraphics[angle=-90,scale=0.52]{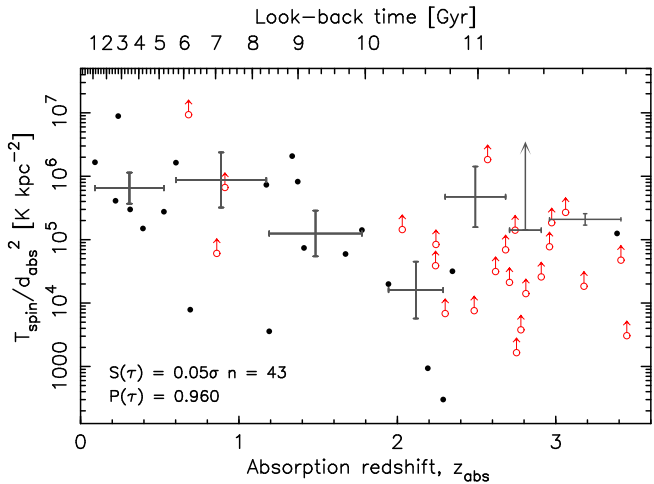}
\caption{The spin temperature degenerate/absorber size versus the redshift.} 
\label{Toverd}
\end{figure}

Referring to Equ. \ref{f}, however, we do know the angular diameter distances and in many cases the extent of the
emission is known from high resolution radio imaging (\citealt{klm+09,kem+12,kps+14,ekp+12}), 
which are resolved at $\theta_{\rm QSO} = 1.6 - 82$ mas. 
So in addition to removing the angular diameter distances (Sect. \ref{cfg}), we can remove $d_{\rm QSO}= \theta_{\rm QSO}
DA_{\rm QSO}$, giving
\begin{equation}
\frac{T_{\rm spin}}{d_{\rm abs}^{~~~2}} = \frac{N_{\rm HI}}{1.823\times10^{18}(\theta_{\rm QSO} DA_{\rm abs})^2\int\!\tau_{\rm obs}\,dv}, 
\label{remove}
\end{equation}
where $\theta_{\rm QSO} DA_{\rm abs}$ is the linear size of the QSO at $z_{\rm abs}$.
Showing this in Fig. \ref{Toverd}, 
we see that any hint of a correlation disappears, although a dip at $z_{\rm abs}\sim2$ persists.
The dip has a value of $\log_{10}(T_{\rm spin}/d_{\rm abs}^{~~~2}) = 4.21\pm0.45$ at $z_{\rm abs} = 2.12\pm0.71$ and 
comparing this with the neighbouring bins,
\begin{itemize}
\item[--]$z_{\rm abs} = 1.48\pm0.29$ is offset by $1.99\sigma$ in $\log_{10}(T_{\rm spin}/d_{\rm abs}^{~~~2})$,
although this bin may be part of the dip, so
\item[--]$z_{\rm abs} = 0.89\pm0.29$ is offset by $3.87\sigma$, 
\item[--]$z_{\rm abs} = 2.49\pm0.19$ is offset by $3.67\sigma$. 
\end{itemize}
Therefore the dip is significant. The bin size is chosen to minimise the remaining unbinned data, while
being sufficiently coarse to display any trends in the data. For example, there are 43 data points in Fig. \ref{Toverd},
which are binned into groups of six leaving one unbinned point (at the highest $z_{\rm abs}$).
If we rebin the data, in three bins of 14 (again with a remainder of one), for
\begin{itemize}
\item[--]$z_{\rm abs} = 0.71\pm0.62$, $\log_{10}(T_{\rm spin}/d_{\rm abs}^{~~~2}) =5.73\pm0.27$,
\item[--]$z_{\rm abs} = 1.97\pm0.60$, $\log_{10}(T_{\rm spin}/d_{\rm abs}^{~~~2}) =4.96\pm0.32$,
\item[--]$z_{\rm abs} = 3.02\pm0.40$, $\log_{10}(T_{\rm spin}/d_{\rm abs}^{~~~2}) =5.35\pm0.07$,
\end{itemize}
and so the dip persists in the middle bin (see Fig. \ref{corr-SFR}). The first bin is offset by $2.40\sigma$ from this, but the 
last is only $1.22\sigma$, although this may be related to the star formation history (see Sect. \ref{sec:SFR}).

 \subsection{Metallicity}

\begin{figure}
\centering \includegraphics[angle=-90,scale=0.52]{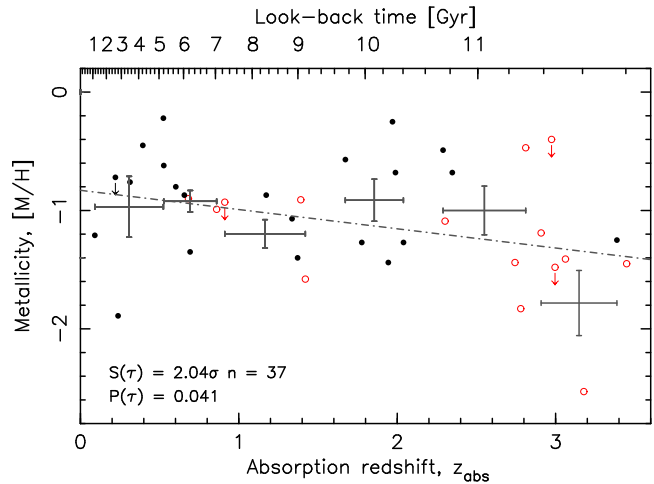}
\caption{The metallicity versus redshift for the 21-cm absorption searches, where available. 
The line shows the least-squares fit to the data.} 
\label{z-M}
\end{figure}

 Since the heavy element abundance is expected to decrease with look-back time
 (e.g. \citealt{pgw+03,cwmc03,rwp+12}),
 in addition to the metals providing radiation pathways for the gas to cool (e.g. \citealt{whm+95}), it has been argued
 that the anti-correlation between the (estimated) spin temperature and metallicity, [M/H], is evidence for an evolution in
 $T_{\rm spin}$ with redshift \citep{ksbc09}.
However, of the 13 low redshift ($z_{\rm abs}\lapp1$) DLAs which have measured metallicities (Fig.~\ref{z-M}),
 only three are non-detected in 21-cm absorption, which does not support the hypothesis of \citet{kc02} that a mix
of spin temperatures should give a mix of metallicities. 

\citet{ctp+07} suggested that the anti-correlation between $T_{\rm spin}/f$ and [M/H] could be driven by the covering factor:
A correlation
between metallicity and velocity spread in low ionisation species had been noted (\citealt{lpf+06,mcw+07}). If
attributed to galactic dynamics \citep{wp98,kkp+07,pcw+07},  this spread indicates that metallicity is a tracer of mass, 
perhaps supporting the argument that the covering factor, through galaxy size, is dominant in the correlation.

\begin{figure}
\centering \includegraphics[angle=-90,scale=0.52]{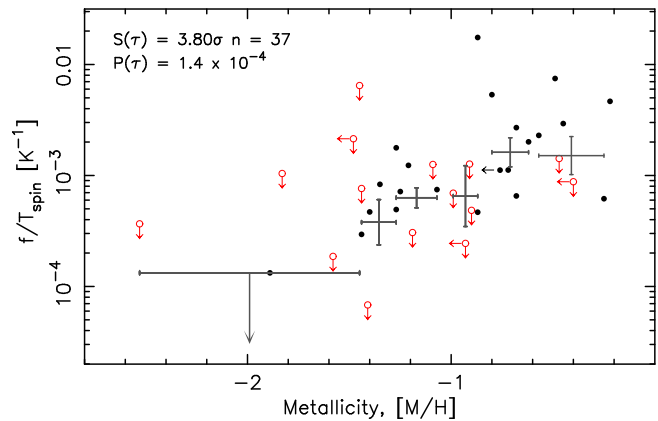}
\caption{$f/T_{\rm spin}$ versus the metallicity. We invert the ordinate from $T_{\rm spin}/f$ since {\sc asurv} cannot mix upper and lower limits.}
\label{foverT-M}
\end{figure}

\begin{figure}
\centering \includegraphics[angle=-90,scale=0.52]{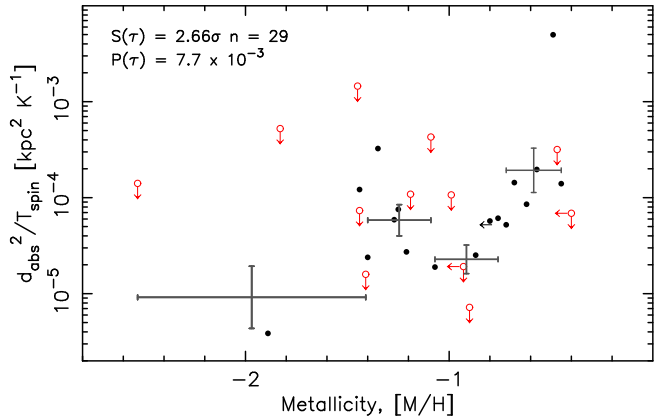}
\caption{As Fig, \ref{foverT-M}, but with the removal of the angular diameter distances and $d_{\rm QSO}$ (cf. Equ. \ref{remove}).}
\label{doverT-M}
\end{figure}

We update the plot of $T_{\rm spin}/f$ versus $z_{\rm abs}$ in Fig. \ref{foverT-M}, 
where we see the correlation strengthens over the previous values of 
$2.80\sigma$ \citep{ctp+07}  and $3.07\sigma$ \citep{ctd+09}. 
Again, however, it is difficult to break the $T_{\rm spin}/f$ degeneracy, although we can apply the above methods to yield $d_{\rm abs}^{~~~2}/T_{\rm spin}$ (Fig. \ref{doverT-M}).
This weakens the correlation, although the nine missing data points, due to not having a measurement of $\theta_{\rm QSO}$, are
mostly detected points (which are used to determine the censored values) from one end of the distribution (top right). 
So a weakening of the correlation is not unexpected and it remains plausible that $d_{\rm abs}\propto$\,[M/H]. 

\section{Star formation history}
\label{sec:SFR}

Although we find no clear evolution in the spin temperature degenerate with absorber size, the binned values of 
$T_{\rm spin}/d_{\rm abs}^{~~~2}$ are indicative of a dip at $z_{\rm abs} \approx2$ (Fig.~\ref{Toverd}). This is close
to where the star formation rate density, 
$\psi_{*}$, peaks \citep{hb06,bbg+13,ssb+13,lbz+14,md14,zjd+14}.  
In contrast, the mass density of neutral hydrogen rises from  $\Omega_{\rm HI} \approx0.5\times10^{-3}$ at $z\lapp0.5$
\citep{zvb+05,lcb+07,bra12,dsmb13,rzb+13,hsf+15,npr+16}, but remains flat at $\Omega_{\rm HI} \approx1\times10^{-3}$ over
$0.5 \gapp z \gapp 5$ \citep{rt00,ph04,rtn05,cur09a,pw09,npc+12,cmp+15}. This has led to a much debated disparity
between the fuel for star formation and the star formation history.

\begin{figure}
\centering \includegraphics[angle=-90,scale=0.55]{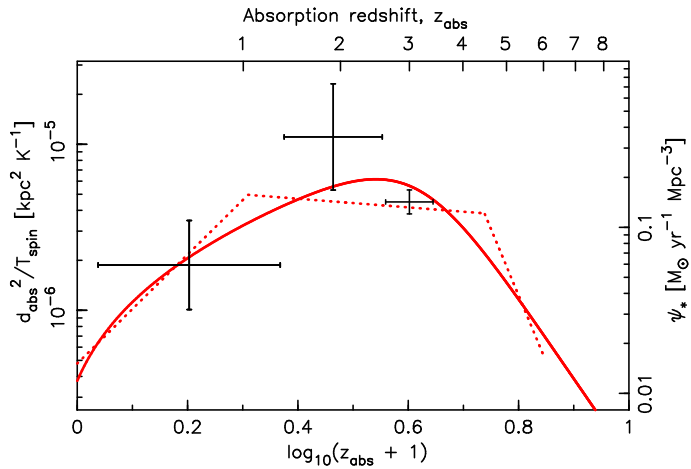}
\caption{The error bars in Fig. \ref{Toverd} with the ordinate re-binned and inverted and the abscissa mapped to
  $\log_{10}(z_{\rm abs} + 1)$, in order to overlay the best fit curve to the SFR density--- redshift distribution from
  \citet{hb06} [the dotted lines show the piecewise linear fitting results]. These have been arbitrarily shifted on the
  ordinate but retain the relative scaling (right hand scale).}
\label{corr-SFR}
\end{figure}

From the binned values of $d_{\rm abs}^{~~~2}/T_{\rm spin}$,
however, we find a reasonable, albeit coarse,  trace of the star formation history (Fig.~\ref{corr-SFR}).
Given that star formation occurs in cold ($\sim10$ K), dense ($\sim10^3$ atoms \ccm) gas, 
this could indicate that the fraction of the cold neutral medium (CNM, where $T\sim150$ K and $n\sim10$ \ccm)
is a more suitable tracer of the reservoir for star formation than the bulk of the neutral gas, which also consists of the
warm neutral medium (WNM, where $T\sim8000$ K and $n\sim0.2$ \ccm, \citealt{whm+95}). 
By definition, the CNM fraction is ratio of cold to total neutral gas, i.e. $\propto \int\!\tau_{\rm obs}\,dv/N_{\rm HI}$, 
which we see to peak at a similar redshift as $\psi_{*}$, when the geometry effects have been accounted for  (Figs. \ref{DLA-corr} \& \ref{Toverd}).

\section{Caveats}

\subsection{Absorber--QSO alignment}
\label{aqa}

\begin{figure}
\centering \includegraphics[angle=-90,scale=0.52]{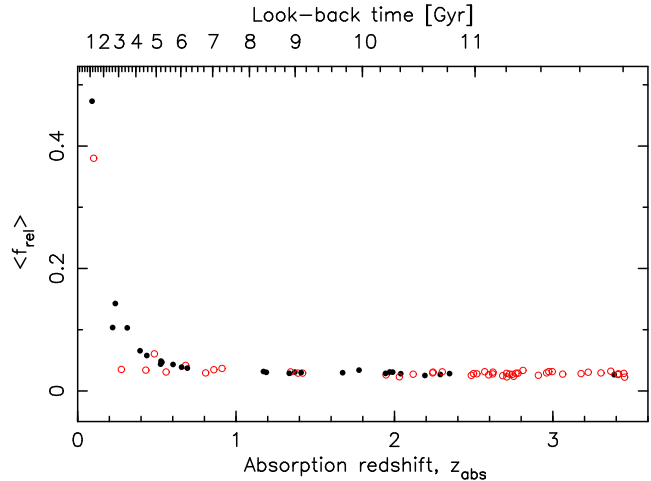}
\caption{The expectation value of the relative covering factor.}
\label{ang_red}
\end{figure}

From Fig. \ref{dla_schem} it is clear that Equ. \ref{f} only holds for direct alignment between the absorber and
the QSO, although, for a given $d_{\rm abs}$ and $d_{\rm QSO}$, interception of the QSO is more likely when
$z_{\rm abs} \ll z_{\rm QSO}$. For a given sight-line, if there is no information on the impact
parameter, the covering factor will vary randomly over $f=0 -1$. We therefore run a 
Monte-Carlo simulation with 10\,000 iterations, where the probability of interception is a random value
between 0 and 1 scaled by the angular diameter distance ratio, $(DA_{\rm QSO}/DA_{\rm abs})^2$,
to yield the expectation value of the covering factor (Fig. \ref{ang_red}).
Since we do not know the absolute covering factor, this is scaled according to the maximum measured ratio, i.e. 
$DA_{\rm QSO}/DA_{\rm abs} = 4.09$, giving  $\left<f_{\rm rel}\right>\approx0.5$ for the lowest redshift point, where the
ratio is normalised out. The distribution essentially traces the angular diameter distance ratios of the sample
\citep{cur12}, but halved and inverted. It should be noted that this is just a statistical relative expectation
value for the covering factor. It is, however,  consistent with the simulations of  \citet{kps+14} which find that the 
covering factors do not vary significantly at $z_{\rm abs} >1$ (cf. Fig. \ref{ang_red}).

\subsection{Absorber sizes}
\label{sec:as}

\begin{figure}
\centering \includegraphics[angle=-90,scale=0.52]{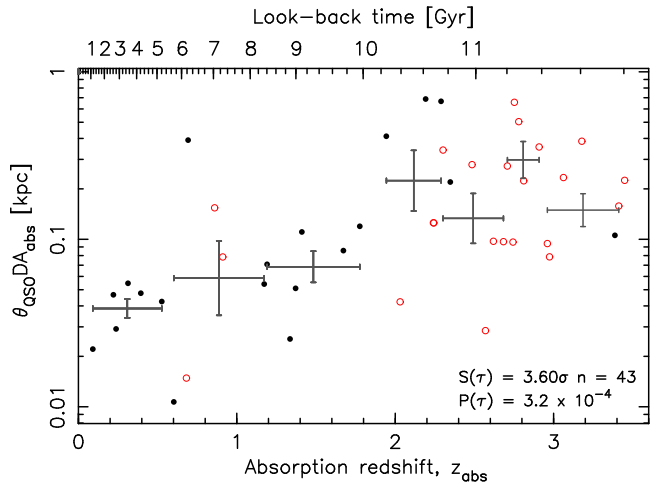}
\caption{The linear extent of the QSO at the absorption redshift versus the absorption redshift.}
\label{abs_size}
\end{figure}

\begin{figure}
\centering \includegraphics[angle=-90,scale=0.52]{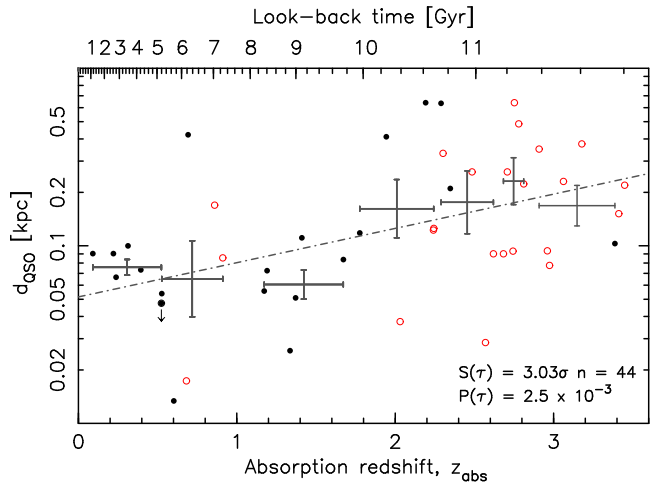}
\caption{The linear extent of the QSO versus the absorption redshift. The line shows the least-squares fit to the data. 
A similar correlation between $d_{\rm QSO}$ and $z_{\rm abs}$ was noted for the 25 measurements then available by  \citet{klm+09}.}
\label{size-z}
\end{figure}

Again, referring to Equ. \ref{f}, the above analysis only holds for $\theta_{\rm abs} < \theta_{\rm QSO}$, i.e. $d_{\rm abs} <
DA_{\rm abs}\theta_{\rm QSO}$, otherwise $f=1$ and the spin temperature can be
determined. 
In Fig.~\ref{abs_size}
we show the distribution of $DA_{\rm abs}\theta_{\rm QSO}$ with $z_{\rm abs}$ for which we see that the lowest redshift
absorbers need a projected extent exceeding $d_{\rm abs} \sim50$ pc to have $f\approx0.5$, taking into account the
alignment (Sect. \ref{aqa}), requiring $d_{\rm HI} \gapp0.1$ kpc for $f\approx1$ (Fig. \ref{dla_schem}). However, without knowing 
the projected size, $d_{\rm abs}$, 
we cannot say which of the sample are sufficiently large to have a covering factor of unity, although the deprojected 
sizes are suspected
to be $d_{\rm HI} \sim0.1-1 $ kpc in the case of near-by associated absorption systems \citep{bra12,cag+13}. 
If, for the sake of argument, we say that $f\approx1$ for the cases where $DA_{\rm abs}\theta_{\rm QSO}\lapp0.1$ kpc,
this would imply that $f\sim1$ at $z_{\rm abs}\lapp2$, at least where the source sizes have been measured. This, in conjunction
with the flattening of the angular diameter distance above these redshifts (Sect. \ref{cfg}), may indicate that
no correction for geometry is required. However, it is clear from Fig. \ref{abs_size} that the 
high redshift absorbers need to be significantly larger than their low redshift counterparts in order that $f=1$
and therefore $f$ is likely to decrease with redshift. This increase in QSO size is most
likely due to the Malmquist bias favouring the largest and brightest radio sources at high redshift (Fig. \ref{size-z}), 

It is, of course, possible that the increase in radio source size with redshift is matched with an increase in absorber size, but
given that there is a well documented evolution in galaxy sizes, where large galaxies dominate the low redshift 
population \citep{bmce00}
and dwarf galaxies the high redshift population (\citealt{lf03}), this is unlikely.

On this matter, it is possible that the decrease in $d_{\rm abs}^{~~~2}/T_{\rm spin}$ at high redshift (Fig. \ref{corr-SFR}) is due to galaxy size evolution.
For massive ($\geq10^{11}$ \Mo) galaxies, which are the easiest to resolve at high redshift, the average galaxy size
decreases by a factor of $\approx3$ over $0\lapp z\lapp3$ \citep{bib+04,fdg+04,tfn+06,btc+08}. If this was applicable to
DLAs, it would imply $d_{\rm abs}^{~~~2} (z_{\rm abs} =3) \sim0.1 d_{\rm abs}^{~~~2} (z_{\rm abs} =0)$, which could
account for the observed $d_{\rm abs}^{~~~2}/T_{\rm spin}$ distribution, {\em if} this ranged from $z_{\rm
  abs}\sim0$. However, the decrease in $d_{\rm abs}^{~~~2}/T_{\rm spin}$ is only observed from $z_{\rm abs}\gapp2$ and so the evolution of galaxy sizes
cannot account for the increase in $d_{\rm abs}^{~~~2}/T_{\rm spin}$ at $z_{\rm abs}\lapp2$.

\subsection{QSO structure}
\label{qstruc}

Another factor, which is essentially the absorber--QSO alignment (Sect. \ref{aqa}) for a non-uniform flux distribution,
is that the strength of the absorption will also depend upon any structure in the emission.  Here, in the absence of any
information regarding the alignment, Equ. \ref{f} assumes uniform flux over the high resolution images
(Sect. \ref{bcfg}), which span $\lapp1$ kpc at the absorption redshift (Sect. \ref{sec:as}). The specific line-of-sight
geometry between the absorber, which may also have structure, and the various components in the emitter, will obviously
affect the validity the Equs. \ref{f} and \ref{remove}, although not the $T_{\rm spin}\left(\frac{d_{\rm QSO}}{d_{\rm
      abs}}\right)^2$ values (Fig. \ref{DLA-corr}), which are only corrected for the angular diameter distances, with
the distribution also exhibiting the dip. Although individual values of $T_{\rm spin}/d_{\rm abs}^{~~~2}$ may not be
reliable, statistically we expect these effects to average out, unless there is a further, currently unenvisioned,
effect at play.

\subsection{Limited data}

\begin{figure}
\centering \includegraphics[angle=-90,scale=0.52]{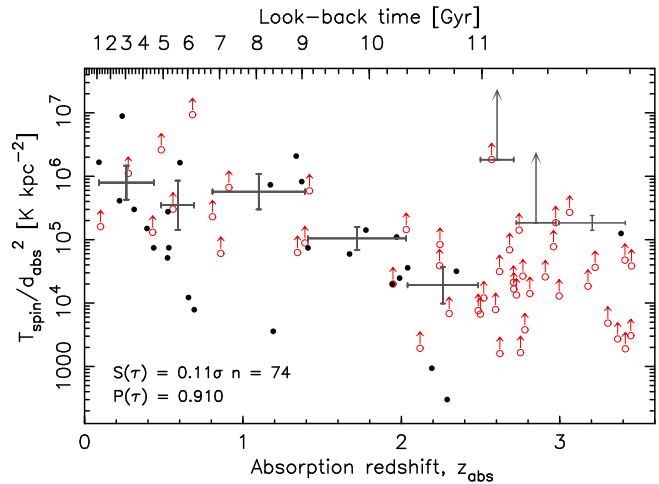}
\caption{As Fig. \ref{Toverd}, but with the unmeasured values of $d_{\rm QSO}$ estimated from the fit in Fig. \ref{size-z}.}
\label{Toverd_est}
\end{figure}

\begin{figure}
\centering \includegraphics[angle=-90,scale=0.55]{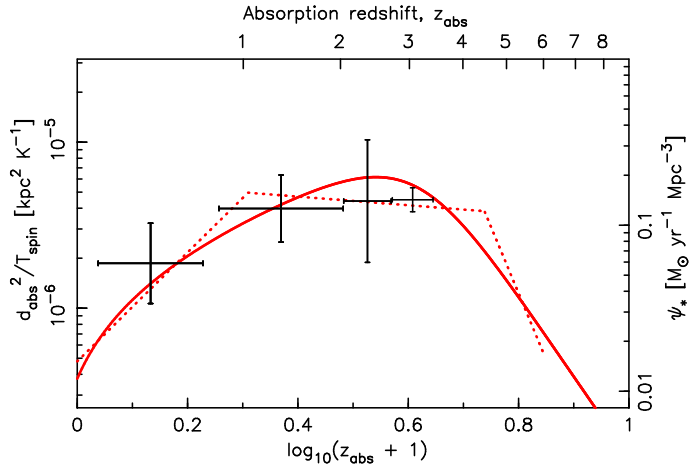}
\caption{As Fig. \ref{corr-SFR}, but with the unmeasured values of $d_{\rm QSO}$ estimated from the fit in Fig. \ref{size-z}. The same ordinate shift is used.}
\label{corr-SFR_est}
\end{figure}

A major caveat of the current analysis is the limited data; only 43 values when we incorporate $d_{\rm QSO}$ (Fig. \ref{Toverd}).
Using the fit in Fig.~\ref{size-z}, we can estimate a typical source size from the correlation with redshift, where this has not been
measured. Showing this in Figs. \ref{Toverd_est} and \ref{corr-SFR_est},
we see that both the $z_{\rm abs}\approx2$ dip and trace of the SFR density persist. 
However, the fact remains that high resolution radio imaging of the current objects, in addition to 21-cm surveys of
$z_{\rm abs} \gapp4$ DLAs, is required 
in order to fully investigate any similarity between $d_{\rm abs}^{~~~2}/T_{\rm spin}$  and the star formation history.

\section{Conclusions}

We investigate the recent assertions that there is an  increase in the spin temperature of damped Lyman-\AL\ absorption
systems with redshift. In order to address this in a comprehensive and self consistent way, we normalise the limits of
the 21-cm absorption searches and include these via a survival analysis, in addition to accounting for the geometry
effects of an expanding Universe. These effects are neglected by other studies but are crucial in any complete
treatment, since they introduce a systematic difference in the covering factor 
between the low and high redshift regimes.

Accounting for the geometry effects, we find the correlation of $T_{\rm spin}/f$ with redshift to 
hold only for $z_{\rm abs} \gapp1$, with an anti-correlation below these redshifts. This indicates
a minimum in the spin temperature (or a maximum in the covering factor), at $z_{\rm abs}\approx 1-2$.
Combining the measurements of the background radio sources with the other {\em known} quantities, the angular diameter
distances to the absorber and the QSO, we can obtain $T_{\rm spin}/d_{\rm abs}^{~~~2}$, where $d_{\rm abs}$ is the
projected diameter of the absorber. Using this, thus avoiding the unjustified assumption that the angular size of the
absorber is equal to that of the compact flux component, we find a dip to persist, with a minimum 
$T_{\rm spin}/d_{\rm abs}^{~~~2}$ at  $z_{\rm abs}\sim2$. Inverting this, giving the fraction of the cold neutral medium, we find
$d_{\rm abs}^{~~~2}/T_{\rm spin}\propto \int\!\tau_{\rm obs}\,dv/N_{\rm HI}$ to follow the star formation history. 

This result does, however, rely upon several assumptions:
\begin{enumerate}
\item That the unknown alignment of the absorber with respect to the sight-line to the QSO, averages out over the sample,
  making the ratio of angular diameter distances the dominant effect. If this is the case, it should also apply where the
   flux is not uniform across the radio emission. 
\item That there is no overwhelming evolution the unknown projected absorber extent, $d_{\rm abs}$, with redshift.
\item That $d_{\rm abs}$ is smaller than the projected extent of the quasar at the absorption redshift. Even if this
is not the case, the covering factor, on the whole, decreases with redshift, requiring a corresponding decrease in spin
temperature to produce the observed $T_{\rm spin}/f-z_{\rm abs}$ distribution.
\end{enumerate}
The result also replies upon the coarse binning of the limited data (43 cases where high resolution images are
available), to average out these effects. However, if new data confirm this result, it would explain why the evolution
of the total neutral hydrogen column density does not trace that of the star formation history and that future efforts
should be focused on the cool component of the gas.

\section*{Acknowledgements}

I wish to thank the referee for their prompt and helpful comments. Also Christian Henkel and James Allison for their
very useful feedback, as well as Warren Trotman (RIP) who {\sf xfig}ed the radio telescope in Fig. \ref{dla_schem}, back
in our Masters days at Jodrell bank.  I never got to thank him in person.  This research has made use of the NASA/IPAC
Extragalactic Database (NED) which is operated by the Jet Propulsion Laboratory, California Institute of Technology,
under contract with the National Aeronautics and Space Administration and NASA's Astrophysics Data System Bibliographic
Service. This research has also made use of NASA's Astrophysics Data System Bibliographic Service and {\sc asurv} Rev
1.2 \citep{lif92a}, which implements the methods presented in \citet{ifn86}.


\label{lastpage}

\end{document}